\documentclass[11pt,a4paper]{article}
\pdfoutput=1 
\usepackage{jheppub}
\usepackage{graphicx}
\usepackage{bm}
\usepackage{amsmath}
\usepackage{amsfonts}
\usepackage{braket}
\usepackage{amssymb}
\usepackage{epstopdf}
\usepackage{hyperref}
\usepackage{mathrsfs}
\usepackage{subfigure}
\usepackage{ulem}

\def\lsim{\mathrel{\rlap{\lower4pt\hbox{\hskip1pt$\sim$}}
    \raise1pt\hbox{$<$}}}    
\def\gsim{\mathrel{\rlap{\lower4pt\hbox{\hskip1pt$\sim$}}
    \raise1pt\hbox{$>$}}}                

\preprint{MAN/HEP/2013/05}
\title{Naturalness of Light Neutralino Dark Matter in pMSSM after    
LHC, XENON100 and Planck Data}
\author[a,b]{C\'{e}line B{\oe}hm,}
\author[c]{P. S. Bhupal Dev,}
\author[d,e]{Anupam Mazumdar}
\author[d]{and Ernestas Pukartas}

\affiliation[a]{Institute for Particle Physics Phenomenology, \\
University of Durham, Durham DH1 3LE, United Kingdom}
\affiliation[b]{LAPTh, U. de Savoie, CNRS, BP 110, 74941 Annecy-Le-Vieux, France}
\affiliation[c]{Consortium for Fundamental Physics, School of Physics and Astronomy,\\ 
University of Manchester, Manchester M13 9PL, United Kingdom}
\affiliation[d]{Consortium for Fundamental Physics, \\
Lancaster University, Lancaster LA1 4YB, United Kingdom}
\affiliation[e]{Niels Bohr Institute, Copenhagen University, Blegdamsvej-17, Denmark.}
\emailAdd{c.m.boehm@durham.ac.uk}
\emailAdd{Bhupal.Dev@hep.manchester.ac.uk}
\emailAdd{a.mazumdar@lancaster.ac.uk}
\emailAdd{e.pukartas@lancaster.ac.uk}

\abstract{We examine the possibility of a light (below 46 GeV) 
neutralino dark matter (DM) 
candidate within the 19-parameter phenomenological Minimal Supersymmetric 
Standard Model (pMSSM) in the light of various recent 
experimental results, especially from the LHC, XENON100, and Planck. We also study the extent of electroweak fine-tuning for such a light neutralino scenario in view of the null results from the searches for supersymmetry so far. 
Using a Markov Chain Monte Carlo likelihood analysis of the full pMSSM 
parameter 
space, we find that a neutralino DM with mass $\gsim 10$ GeV can in principle 
still satisfy all the existing constraints. Our light neutralino solutions can be broadly divided into two regions: (i) The solutions in the 10 - 30 GeV 
neutralino mass range are highly fine-tuned and require the existence 
of light selectrons (below 100 GeV) in order to satisfy the observed DM relic 
density. We note that these are not yet conclusively ruled 
out by the existing LEP/LHC results, and a dedicated analysis valid for a 
non-unified gaugino mass spectrum is 
required to exclude this possibility.  
(ii) The solutions with low fine-tuning are mainly in the 30 - 46 GeV 
neutralino mass range. However, a major portion of it is already ruled 
out by the latest XENON100 upper limits on its spin-independent direct 
detection cross section, and the rest of the allowed points are within the XENON1T projected limit. Thus, we show that the allowed MSSM parameter space 
for a light neutralino DM below the LEP limit of 46 GeV, possible in supersymmetric models without gaugino mass 
unification, could be 
completely accessible in near future. This might be useful in view of 
the recent claims for positive hints of a DM signal in some 
direct detection experiments.} 
\begin{document}
\maketitle
\section{Introduction}\label{sec1}
Low-scale Supersymmetry (SUSY) (see e.g.,~\cite{rohini, baerbook}) 
is 
one of the most attractive candidates for New Physics beyond the Standard Model 
(SM). Apart from providing successful gauge coupling unification and a solution to the gauge hierarchy problem, it offers a natural candidate for Dark Matter (DM) in our Universe in the form of the lightest supersymmetric particle (LSP). In the $R$-parity conserving Minimal Supersymmetric Standard Model (MSSM), the lightest neutralino ($\widetilde{\chi}_1^0$) is one of the most viable Weakly Interacting Massive Particle (WIMP) DM candidates (for a review, see e.g.,~\cite{susy-dm1, susy-dm2}). It can explain the observed DM relic density, while predicting experimentally accessible direct and indirect detection rates, over a wide range of supersymmetric model parameters, some of which are already getting constrained from the ongoing direct searches for the supersymmetric particles at the Large Hadron Collider (LHC), in combination with other low-energy data (for a review, 
see e.g.,~\cite{PDG}).    

The recent hints of positive signals in three DM direct detection experiments, namely, DAMA~\cite{dama1, dama2}, CoGeNT~\cite{cogent1} and CRESST-II~\cite{cresst}, have generated a lot of interest in {\it light} WIMP candidates in the 5 - 50 GeV mass range. This interpretation is however challenged by the null results from various other direct detection experiments, most notably the latest XENON100 results~\cite{Aprile:2012nq}, which provide the most stringent upper limits on the spin-independent WIMP-nucleon scattering cross section for $m_\chi>7$ GeV~\footnote{More recently, the TEXONO experiment~\cite{texono} has achieved slightly better 
sensitivity than XENON100 for $m_\chi<7$ GeV.}. Nonetheless, due to the relatively poor sensitivity of the XENON100 experiment in the very low WIMP mass regime, it is believed that an agreement between the positive and null sets of experimental results could be possible, if at all, only in this low mass region. Hence, it might be worthwhile examining the allowed MSSM parameter space to see if there exists a {\it lower} bound on the lightest neutralino mass irrespective of the direct detection results. 

The neutralino mass eigenstates in the MSSM result from mixing of the neutral bino ($\widetilde{B}$), wino ($\widetilde{W}^0$) and higgsinos ($\widetilde{H}_d^0,\widetilde{H}_u^0$). This mixing is determined by the MSSM $\tan\beta$ parameter and the bino, wino and higgsino mass parameters $M_1,M_2$ and $\mu$, respectively. In the SUSY models with gaugino mass unification at the Grand Unified Theory (GUT) scale, a relation between the bino and wino mass follows at the electroweak scale:  $M_1=\frac{5}{3}\tan^2 \theta_W M_2\approx 0.5 M_2$ (see e.g.,~\cite{Martin:1997ns}) which, after mixing, translates into a chargino-neutralino mass relation. Therefore, a lower limit on the lightest neutralino ($\widetilde{\chi}^0_1$) mass of about 46 GeV 
can be derived for these models from the Large Electron Positron (LEP) chargino mass limit~\cite{LEP-susy}, whereas in the constrained MSSM (cMSSM)~\cite{cMSSM} with both gaugino and sfermion mass unification, this limit increases to well above 100 GeV from the strong constraints set by the recent LHC data~\cite{PDG}.  

On the other hand, in a generic MSSM scenario without the assumption of gaugino 
mass unification, there is no general lower limit on the lightest neutralino mass~\cite{Dreiner:2009ic}. The LEP limit on the invisible decay width of the SM $Z$ boson applies to light neutralinos with a mass below $m_Z/2=45.6$ GeV, but it depends on the $Z\widetilde{\chi}_1^0\overline{\widetilde{\chi}}_1^0$ coupling which could be small or even zero, depending on the higgsino component of the neutralino. In such a 
case, light neutralinos are mainly constrained by the DM relic density 
measurement as 
well as by the collider and flavor constraints on the SUSY parameter space. 
Assuming that the lightest neutralino is non-relativistic and 
provides the entire cold DM content of 
the Universe, while satisfying the LEP bounds on chargino and stau masses, 
Ref.~\cite{hooper} obtained a lower bound of $m_{\widetilde{\chi}^0_1}\gsim 
18$ GeV. This was relaxed to about 6 GeV without violating the LEP bounds and flavor sector constraints in SUSY models with a pseudo-scalar Higgs boson ($A$) mass $m_A<200$ GeV and a large $\tan\beta$~\cite{bottino, belanger03}. This was even further lowered to about 3 GeV by allowing explicit $CP$ violation in the MSSM Higgs sector~\cite{scopel}.     
      
Meanwhile, several new experimental results have been obtained at the LHC: (i) A new Higgs-like neutral scalar particle has been discovered~\cite{atlas-higgs,cms-higgs} with mass around 125 GeV which falls squarely within the 
MSSM-predicted range for the lightest $CP$-even neutral Higgs mass: $m_h \in [115,135]$ GeV (see e.g.,~\cite{djouadi-review}); (ii) The rare decay $B_s^0\to \mu^+\mu^-$ was observed with a branching fraction of ${\cal B}(B_s^0\to \mu^+\mu^-)=\left(3.2^{+1.5}_{-1.2}\right)\times  10^{-9}$~\cite{lhcb} which is in agreement with the SM expectation, ${\cal B}(B_s^0\to \mu^+\mu^-)_{\rm SM}=(3.23\pm 0.27)\times  10^{-9}$~\cite{bsmu-SM}; (iii)  The lack of a SUSY signal at the $\sqrt{s}=7$ and 8 TeV LHC has pushed the lower limits on the squark and gluino masses to about 1 TeV and beyond~\cite{atlas-susy,cms-susy}; (iv) Updated bounds have been obtained for the MSSM Higgs sector~\cite{atlas-susy-higgs1, atlas-susy-higgs2, cms-susy-higgs}. All these new results have 
profound implications for a light neutralino scenario within the MSSM, and some of these aspects have already been investigated in a number of recent 
studies 
~\cite{nath, zurek, boehm1, boehm2, fornengo, fornengo0, Calibbi, Choudhury:2012tc, Arbey:2012na, lindner, Boehm:2012rh, Belanger:2012jn,han, Cheung:2012qy, Altmannshofer:2012ks}. The general conclusion is that light neutralino 
DM candidates with mass below about 15 GeV are severely constrained in generic 
MSSM scenarios (without gaugino mass unification).    

Another important issue to be addressed in the light of the recent LHC results is the apparent ``little hierarchy problem", i.e., how does a multi-TeV SUSY particle spectrum conspire to give a weak-scale $Z$-boson mass and also 
a Higgs boson mass of 
125 GeV? One way of analyzing this issue quantitatively is to evaluate the measure of electroweak fine-tuning (EWFT) by examining the minimization condition in the MSSM Higgs potential which determines the $Z$-boson mass~\cite{Ellis:1986yg, ewft}. It is well-known that radiative corrections must play a crucial role in determining the allowed SUSY parameter space necessary to generate a 125 GeV Higgs boson mass, much larger than its tree-level prediction of $m_h\leq m_Z$. This in general can lead to a large fine-tuning. In addition to this, the requirement of a 
light neutralino DM will pose a challenge for any MSSM scenario, the severity of which is however strongly model-dependent. The naturalness of various SUSY models with a neutralino LSP has been analyzed in the literature (for an incomplete list, see~\cite{lindner, King, Horton, Perel1, Arbey:2011a, Arbey:2011b, 
Hewett2, Amsel, Bechtle:2012zk, Hewett, Perelstein:2012qg, Antusch:2012gv, Baer:2012mv, Gogoladze:2012yf, Fichet:2012sn}, and references therein).  

In this paper, we perform a dedicated study focusing on the naturalness of a 
{\it light} neutralino, and also examining how light the neutralino 
could be, after taking into account all the existing theoretical and experimental constraints. To perform such an analysis in the full 124-parameter MSSM is 
quite unrealistic, and hence, we need to make some well-motivated 
assumptions in order to reduce the number of parameters to a manageable level. 
Most of the earlier studies on SUSY focused on the cMSSM having only 5 parameters, assuming certain boundary conditions at the 
GUT-scale.  However, in view of the latest null results from SUSY searches at the $\sqrt s=8$ TeV LHC, the cMSSM seems too restrictive for low-scale SUSY phenomenology as the allowed cMSSM parameter space accessible to the $\sqrt s = 14$ TeV LHC is rapidly shrinking (for the latest global status, see e.g., Ref.~\cite{cMSSMglobal3, cMSSMglobal2, cMSSMglobal1}). Therefore, in this paper we 
choose not to make any assumptions at the high scale, and focus only on the low-scale MSSM parameter space from a phenomenological point of view. 
More precisely, we consider a $CP$-conserving MSSM (i.e., with no new $CP$ phases) with Minimal Flavor Violation~\cite{mfv} and with first two generations of sfermions degenerate. This is widely known as the phenomenological 
MSSM (pMSSM)~\cite{pMSSM} (also known as `SUSY without prejudice'~\cite{Berger:2008cq}) with 19 free parameters at the electroweak scale. We also study the level of fine-tuning for the light neutralino scenario in this 
context. 

In order to efficiently explore the 19-dimensional pMSSM parameter space, we perform a Markov Chain Monte Carlo (MCMC) likelihood analysis (for a review, see e.g.,~\cite{mcmc}), with the priors chosen to focus on a light neutralino 
scenario with mass below the conservative LEP lower limit of 46 GeV. 
We include in our analysis the latest experimental results for SUSY searches from the LHC~\cite{atlas-susy, cms-susy} which now supersede the Tevatron results~\cite{cdf, d0}, in addition to the existing LEP limits~\cite{LEP-susy}, wherever applicable. We also include the latest astrophysical/cosmological constraints for a WIMP DM from the 9-year WMAP data~\cite{wmap} as well as the very recently released Planck data~\cite{Planck}. We further examine the allowed parameter space in the light of various recent results for DM 
direct detection, most notably the XENON100 limits~\cite{Aprile:2012nq}, as well as the indirect detection results from Fermi-LAT data~\cite{Ackermann:2012qk}.

We find that a light neutralino DM with mass as low as 10 GeV is still allowed in the pMSSM, while satisfying all the existing experimental constraints provided we only take the model-independent analysis results from LEP. (Including the 
LEP limits strictly applicable to gaugino-mass unification models allows only the solutions with $m_{\widetilde{\chi}^0_1}>30$ GeV, in agreement with previous results~\cite{boehm1, boehm2}.) However, such neutralinos which are required to be mostly bino-like are severely fine-tuned and require the existence of light sleptons below 100 GeV in order to provide an efficient annihilation channel to reduce the bino relic density to be consistent with the observed limit. A dedicated analysis of the LEP data in the context of a pMSSM scenario could eliminate this region completely. On the other hand, low fine-tuning regions can be obtained around $m_{\widetilde{\chi}^0_1}=45$ GeV where the resonant annihilation via the $s$-channel 
$Z$-exchange is possible for the neutralino with a non-negligible higgsino component. However, such regions are mostly excluded by the recent XENON100 limits on the spin-independent DM-nucleon scattering cross section, 
and the remaining such points are well within the reach of the XENON1T~\cite{Aprile:2012zx} and LUX~\cite{LUX} projected limits.

Our paper is organized as follows: In Section~\ref{sec2}, we briefly discuss the electroweak fine-tuning measure. In Section~\ref{sec3}, we list all the pMSSM parameters and their scan ranges, and also summarize all the relevant experimental constraints used in our numerical analysis. In Section~\ref{sec4}, we present our scan results and discuss their implications for a light neutralino DM. Finally, our conclusions are given in Section~\ref{sec5}.  
\section{Electroweak Fine-Tuning}\label{sec2}
To quantify the amount of fine-tuning in the electroweak symmetry breaking (EWSB) sector of the MSSM, it is sufficient for us to analyze the tree-level MSSM 
scalar potential. Since it allows only charge-conserving vacua, we only have to minimize the scalar potential for the neutral scalar fields~\cite{rohini}:
\begin{eqnarray}
V &=& (m_{H_u}^2+\mu^2)|H_u^0|^2+(m_{H_d}^2+\mu^2)|H_d^0|^2
\nonumber\\ &&
-B\mu(H_u^0H_d^0+{\rm H.c.})+\frac{1}{8}(g^2+g'^2)(|H_u^0|^2-|H_d^0|^2)^2 \, ,  
\label{scalar} 
\end{eqnarray}
where $\mu$ is the SUSY-preserving bilinear Higgs superpotential parameter, 
$m_{H_{u,d}}$ and $B$ are the soft scalar masses and the bilinear coupling in 
the SUSY-breaking sector respectively, and $g,g'$ the $SU(2)_L$ and $U(1)_Y$ 
gauge couplings, respectively. After minimization, we obtain the well-known relation for the $Z$-boson mass:
\begin{eqnarray}
 \label{minim}
\frac{m_Z^2}{2}= \frac{m_{H_d}^2-m_{H_u}^2\tan^2\beta}{\tan^2\beta-1}-\mu^2,
\end{eqnarray}
where $\tan\beta=v_u/v_d$ is the ratio of the vacuum expectation values (vevs) of the two Higgs doublet fields $H_u$ and $H_d$ respectively,  It is clear from Eq.~(\ref{minim}) that a cancellation of the terms on the right hand side (RHS) is required in order to obtain the measured value of $m_Z=91.2$ GeV~\cite{PDG} especially if the mass parameters on the RHS are orders of magnitude larger than the weak scale which indeed seems to be the case, given the current experimental status of the direct SUSY searches~\cite{atlas-susy, cms-susy}. Thus naively speaking, the weak scale value of $|\mu|$ can be used as a measure of fine-tuning in the MSSM. 

A more sophisticated way to quantify the degree of EWFT is by using log-derivatives~\cite{ewft}:
\begin{equation}
\label{deltap}
\Delta p_i=\bigg\lvert\frac{\partial\ln m_Z^2(p_i)}{\partial\ln p_i}\bigg\rvert=\bigg\lvert\frac{p_i}{m_Z^2}\frac{\partial m_Z^2}{\partial p_i}\bigg\rvert \, , 
\end{equation}
where $p_i$'s are the parameters that determine the observable $Z$-mass at tree-level. From Eq.~(\ref{minim}), we have $p_i=\{\mu^2,b,m_{H_u},m_{H_d}\}$ (with $b\equiv B\mu$), and the total measure of the EWFT is defined as
\begin{eqnarray}
\Delta_{\rm tot}=\sqrt{(\Delta\mu^2)^2+(\Delta b)^2+(\Delta m_{H_u}^2)^2+(\Delta m_{H_d}^2)^2},
\label{delta-tot}
\end{eqnarray}
with the individual $\Delta p_i$'s given by~\cite{Perelstein:2007nx}
 \begin{eqnarray}
 \Delta\mu^2 &=& \frac{4\mu^2}{m_Z^2}\bigg(1+\frac{m_A^2+m_Z^2}{m_A^2}
\tan^2 2\beta\bigg),\nonumber\\
\Delta b &=& \bigg(1+\frac{m_A^2}{m_Z^2}\bigg)\tan^2 2\beta,\nonumber\\
 \Delta m_{H_u}^2 &=&\bigg\lvert\frac{1}{2}\cos2\beta+\frac{m_A^2}{m_Z^2}
\cos^2\beta-\frac{\mu^2}{m_Z^2}\bigg\rvert
\bigg(1-\frac{1}{\cos2\beta}+
 \frac{m_A^2+m_Z^2}{m_A^2}\tan^2 2\beta\bigg),\nonumber\\
 \Delta m_{H_d}^2 &=&\bigg\lvert-\frac{1}{2}\cos2\beta+
\frac{m_A^2}{m_Z^2}\sin^2 \beta-\frac{\mu^2}{m_Z^2}\bigg\rvert 
\bigg(1+\frac{1}{\cos2\beta}+
 \frac{m_A^2+m_Z^2}{m_A^2}\tan^2 2\beta\bigg).
\label{delta}
\end{eqnarray}
Here $m_A^2=B\mu(\tan\beta+\cot\beta)$ is the MSSM pseudo-scalar Higgs mass, 
and we have assumed $\tan\beta>0$.  Values of $\Delta_{\rm tot}\gg 1$ indicate 
significant fine-tuning. Note that in the decoupling limit with $m_A\gg m_Z$ and with large $\tan\beta$, the 
quantities $\Delta m_{H_u}^2$ and $\Delta m_{H_d}^2$ in Eq.~(\ref{delta}) are 
small, and 
\begin{eqnarray}
\Delta \mu^2 \simeq \frac{4\mu^2}{m_Z^2} \, , \quad \Delta b \simeq \frac{4m_A^2}{m_Z^2 
\tan^2\beta} \, .
\end{eqnarray}
Thus in the limit of large $\tan\beta$, we recover the naive result that fine-tuning increases with increasing $|\mu|$. 

We should note here that including loop corrections to Eq.~(\ref{minim}), one finds the largest contribution to be coming from the (s)top loop which feeds into 
the 
soft mass of the Higgs~\cite{Martin:1997ns}:
\begin{eqnarray}
\delta m_{H_u}^2 = -\frac{3y_t^2}{8\pi^2}(m^2_{\widetilde{t}_L}+m^2_{\widetilde{t}_R}+|A_t|^2)\log\left(\frac{\Lambda}{m_{\widetilde{t}}}\right),
\label{loop-mHu}
\end{eqnarray} 
where $y_t$ is the top-quark Yukawa coupling, $A_t$ is the third-generation 
$A$-term in the SUSY-breaking sector, and $\Lambda$ is some high scale where the stop masses $m_{\widetilde{t}_{L,R}}$ are generated from the SUSY-breaking mechanism. Even for a low-scale SUSY-breaking scenario (such as gauge mediation), this requires a fine-tuning of at least a few percent in order to get the observed $Z$-mass and it becomes worse for heavy stop masses
~\cite{Chang:2005ht}. A heavy physical stop mass $m_{\widetilde{t}}\gsim 500$ GeV is anyway required to provide large enough radiative corrections to the light $CP$-even Higgs mass to raise it to the vicinity of 125 GeV from its tree-level value $\leq m_Z$~\cite{Martin:1997ns}:
\begin{eqnarray}
\delta m^2_{h^0} &=& \frac{3m_t^4}{16\pi^2 v^2}\left[\log\left(\frac{m^2_{\widetilde{t}_L}m^2_{\widetilde{t}_R}}{m_t^4}\right)
+\frac{X_t^2}{m_{\widetilde{t}_L}m_{\widetilde{t}_R}}\left(1-\frac{X_t^2}{12m_{\widetilde{t}_L}m_{\widetilde{t}_R}}\right)\right]\, ,
\label{deltamh}
\end{eqnarray}
 where $X_t=A_t-\mu\cot\beta$ is the stop mixing parameter. A fine-tuning measure $\Delta_{m_h}$ for the Higgs mass can be defined analogous to Eq.~(\ref{deltap}), and for a particular choice of some of the SUSY parameters, it was found 
that $\Delta_{m_h}>75~(100)$ in order to achieve a Higgs mass of 124 (126) GeV and the corresponding stop mass is always heavier than 300 (500) GeV~\cite{Hall:2011aa}. The fine-tuning due to Eqs.~(\ref{loop-mHu}) and (\ref{deltamh}) could 
in principle be added in quadratures to our ``total" fine-tuning parameter defined 
by Eq.~(\ref{delta-tot}), but since their specific values are scale-dependent~\footnote{The SUSY-scale by convention is usually taken to be the geometric mean 
of the two stop masses.} and adds to some arbitrariness in its definition, we do not include them in our analysis.  

\section{Parameters and Constraints}\label{sec3}
We consider the pMSSM with 19 free parameters at the SUSY scale, 
as shown in Table~\ref{table1}. We further assume that the lightest neutralino is the LSP, and our goal is to examine how light and natural the neutralino could be while satisfying all the existing theoretical and experimental constraints on the MSSM parameter space. In particular, we focus on the lightest neutralino masses below the LEP limit of 46 GeV~\cite{LEP-susy} which is strictly  
valid assuming gaugino mass unification at the GUT-scale. 
We perform a numerical scan over the 
19-dimensional pMSSM parameter space using a MCMC-based likelihood analysis for  a light neutralino with the prior ranges given in Table~\ref{table1}.  
These particular ranges are chosen in order to economize the scan time and to 
focus only on the SUSY parameter space not yet disfavored 
by the combined direct search results from the LEP~\cite{LEP-susy}, Tevatron~\cite{cdf,d0} and 
LHC~\cite{atlas-susy, cms-susy}, as discussed later in this section. For the SM parameters $\alpha_s(m_Z),~\alpha_{\rm em} (m_Z),~m_W,~m_t$ and $m_b$, we use the standard values as given in Ref.~\cite{PDG}. 
\begin{table}[ht!]
\begin{center}
\begin{tabular}{c|c|c}
\hline\hline 
  Parameter & Description & Prior Range \\ 
\hline  
       $\tan\beta$ & Ratio of the scalar doublet vevs & [1, 60] \\ 
       $\mu$ & Higgs-Higgsino mass parameter &  [$-3$, 3] TeV \\ 
       $M_A$ & Pseudo-scalar Higgs mass & [0.3, 3] TeV   \\ 
       $M_1$  & Bino mass & [$-0.5$, 0.5] TeV \\ 
       $M_2$ & Wino mass & [$-1$, 1] TeV  \\ 
       $M_3$ & Gluino mass & [0.8, 3] TeV \\ 
       $m_{\widetilde{q}_{L}}$ & First/second generation $Q_L$ squark &  [0, 3] TeV \\  
       $m_{\widetilde{u}_{R}}$ & First/second generation $U_R$ squark & [0, 3] TeV \\  
       $m_{\widetilde{d}_{R}}$ & First/second generation $D_R$ squark & [0, 3] TeV \\  
       $m_{\widetilde{\ell}_{L}}$ & First/second generation $L_L$ slepton & [0, 3] TeV  \\  
       $m_{\widetilde{e}_{R}}$ & First/second generation $E_R$ slepton & [0, 3] TeV \\  
      $m_{\widetilde{Q}_{3L}}$ & Third generation $Q_L$ squark &  [0, 3] TeV \\  
     $m_{\widetilde{t}_{R}}$ & Third generation $U_R$ squark & [0, 3] TeV \\  
       $m_{\widetilde{b}_{R}}$ & Third generation $D_R$ squark & [0, 3] TeV \\ 
       $m_{\widetilde{L}_{3L}}$ & Third generation $L_L$ slepton & [0, 3] TeV \\ 
       $m_{\widetilde{\tau}_{R}}$ & Third generation $E_R$ slepton &[0, 3] TeV \\ 
          $A_t$ & Trilinear coupling for top quark  &[$-10$, 10] TeV\\ 
       $A_b$ & Trilinear coupling for bottom quark  &[$-10$, 10] TeV \\ 
       $A_\tau$ & Trilinear coupling for $\tau$-lepton& [$-10$, 10] TeV \\ \hline\hline
\end{tabular}
\end{center}
\caption{The pMSSM parameters and their range of values used in our numerical analysis.}
\label{table1}
\end{table}

Apart from the direct collider constraints on the sparticle masses, there exist various theoretical and experimental constraints which must be imposed on the 
pMSSM parameter space in our analysis. As a standard theoretical requirement, 
our sparticle spectrum for each allowed point in the parameter space must be 
tachyon-free and should not lead to color- and charge-breaking minima in the scalar potential~\cite{Casas:1995pd}. We also require 
that the scalar potential is bounded from below and is consistent with electroweak symmetry breaking. From the radiative electroweak symmetry breaking arguments, we can restrict the $\tan\beta$ parameter to be roughly between 1 - 60, as given in Table~\ref{table1}.  

The various experimental constraints from direct collider searches, Higgs and flavor sectors, and astrophysical/cosmological data used in our analysis are summarized below. 
\subsection*{Invisible $Z$-decay Width } 
The precise measurement of the $Z$-boson decay width at LEP: $\Gamma_Z^{\rm tot}=2495.2\pm 2.3$ MeV~\cite{lepewwg} puts severe constraints on light neutralinos and charginos with mass $<m_Z/2$. From the LEP measurements of the invisible decay width of the $Z$-boson: $\Gamma_Z^{\rm inv} = (499.0\pm 1.5)$ MeV~\cite{lepewwg}, the parameter space for the lightest neutralino in our case is restricted to 
mostly gaugino-like scenarios ($|\mu|\gg M_{1,2}$) since the neutralino coupling to $Z$ is only via its higgsino component. The allowed fraction of the higgsino component for a given neutralino mass can be calculated using the following expression for the partial decay width of the $Z$-boson to neutralinos~\cite{Haber}:
\begin{eqnarray}
\frac{\Gamma(Z\to \widetilde{\chi}^0_1\widetilde{\chi}^0_1)}{\Gamma(Z\to \nu\bar\nu)} = 2\left(1-\frac{4m_{\widetilde{\chi}^0_1}^2}{m_Z^2}\right)^{1/2}\left[\left(1-\frac{m_{\widetilde{\chi}^0_1}^2}{m_Z^2}\right)\left[(O^L_{11})^2+(O^R_{11})^2\right]+\frac{6m_{\widetilde{\chi}^0_1}^2}{m_Z^2}O^L_{11}O^R_{11}\right]
\label{zinv}
\end{eqnarray}
where $\Gamma(Z\to \nu\bar\nu)=(501.62\pm 0.10)$ MeV is the SM contribution to its invisible decay width (for 3 neutrino species). The components 
$O^{L,R}_{ij}$ are defined as  
\begin{eqnarray}
O^L_{ij} = -\frac{1}{2}N_{i3}N^*_{j3}+\frac{1}{2}N_{i4}N^*_{j4}\, , \quad 
O^R_{ij} = -\left(O^L_{ij}\right)^*
\end{eqnarray}
with $N_{ij}$ measuring the gaugino-Higgsino mixing: 
\begin{eqnarray}
\widetilde{\chi}^0_i = \sum_{k=1}^4 N_{ik}\widetilde{\psi}_k^0,\quad
{\rm where}~\widetilde{\psi}^0 = (\widetilde{B},\widetilde{W}^0,\widetilde{H}_u^0,\widetilde{H}_d^0). 
\end{eqnarray}

Using the LEP measurement of the invisible $Z$-decay width, the following 
constraint can be derived:
\begin{eqnarray}
\Gamma(Z\to \widetilde{\chi}^0_i\widetilde{\chi}^0_j) < 3~{\rm MeV} \quad 
{\rm if}~ (m_{\chi_i}+m_{\chi_j})<m_Z
\label{gz}
\end{eqnarray}
Note that this constraint should apply to {\it all} light neutralinos satisfying this condition, not just the LSP ($i=j=1$ in Eq.~\ref{gz}). However, it is unlikely that decays such as $Z\to \widetilde{\chi}^0_2\widetilde{\chi}^0_1$ will be kinematically allowed, and in such cases, $\widetilde{\chi}^0_2$ will mostly decay to visible final states. Similarly, the decays $Z\to \widetilde{\nu}_i\widetilde{\nu}_j$ are not kinematically allowed for the parameter space examined 
here, and hence, they do not contribute to the purely invisible width of the $Z$-boson.
\subsection*{Exclusion Limits from Collider Searches} 
The experimental lower limits on the sparticle masses are usually quoted assuming gaugino and/or sfermion mass universality at the GUT scale. In a generic MSSM setup, most of these constraints can be relaxed, or even circumvented, for example in case of small mass 
splitting with the LSP or in case of small couplings to the SM vector bosons. 
Since we are interested in light neutralino LSPs here, we must carefully interpret the direct search limits in order to be able to include all the allowed pMSSM parameter space.    

{\bf Neutralino}: As already mentioned in Section~\ref{sec1}, there are no rigorous lower limits on the neutralino masses in the MSSM from direct collider 
searches. The LEP limits~\cite{LEP-susy, neutralino1}
\begin{eqnarray}
m_{\widetilde{\chi}_1^0}>46~{\rm GeV} ,   
m_{\widetilde{\chi}_2^0}>62.4~{\rm GeV} ,   
m_{\widetilde{\chi}_3^0}>99.9~{\rm GeV} ,   
m_{\widetilde{\chi}_4^0}>116.0~{\rm GeV}\, 
\end{eqnarray}   
were derived assuming gaugino mass unification at the GUT scale, and hence, relating the neutralino mass to the chargino mass. Moreover, for a (mostly) 
bino-like 
neutralino, which is required to be the case for $m_{\widetilde{\chi}^0_1}<m_Z/2$ in order to avoid the $Z$-width constraint, its production via $s$-channel exchange of $Z/\gamma^*$ is (negligible) absent. The $t$-channel production cross-section via selectron exchange is also expected to be small for selectron 
masses above the LEP limit (see below). Thus, we can easily satisfy the LEP 
upper limits on the 
neutralino pair-production cross sections 
$\sigma(e^+e^-\to \widetilde{\chi}^0_i\widetilde{\chi}^0_j)$~\cite{neutralino1, neutralino2} for a mostly bino-like neutralino LSP. Similarly, the Tevatron~\cite{tev-chi} and LHC~\cite{cms-chi} SUSY searches for final states involving 
$Z$-bosons cannot constrain a bino-like neutralino.   

{\bf Chargino}: Charginos can be pair-produced at LEP via $s$-channel exchange of $Z/\gamma^*$ or $t$-channel exchange of electron-sneutrino, with destructive interference. It dominantly decays to $\ell{\widetilde{\nu}}$, if kinematically allowed. If not, the three-body decay to $f\bar{f}'\widetilde{\chi}^0_i$ via virtual $W$-boson or sfermions becomes important in which case, the final state fermions ($f,\bar{f}'$) are dominantly leptonic (hadronic) if the sleptons are light (heavy). From the combined searches in fully-hadronic, semi-leptonic and 
fully-leptonic decay modes, LEP has derived a general lower limit of 103.5 GeV~\cite{LEP-susy} which is valid for pMSSM as well, except in corners of phase space where (i) the detection 
efficiencies are reduced, e.g., when the mass differences $\Delta m_+ =m_{\widetilde{\chi}^\pm_1}-m_{\widetilde{\chi}^0_1}$ or $\Delta m_\nu =m_{\widetilde{\chi}^\pm_1}-m_{\widetilde{\nu}}$ are very small (below a few GeV); or (ii) the chargino production cross section is suppressed, e.g., when the electron sneutrino mass is small, thus leading to a destructive interference between $s$- and $t$-channel Feynman diagrams. Dedicated searches for such scenarios have also been performed. For instance, for small $\Delta m_+<3$ GeV but with large sneutrino 
mass, the limit becomes $m_{\widetilde{\chi}^\pm_1}>91.9$ GeV for degeneracy in the gaugino region ($|M_1|\sim |M_2|\ll |\mu|$) while $m_{\widetilde{\chi}^\pm_1}>92.4$ GeV for degeneracy in the higgsino region ($|\mu|\ll |M_1|,|M_2|$)~\cite{LEP-susy}. Without assuming gaugino mass unification, a lower limit of 
$m_{\widetilde{\chi}^\pm_1}>70$ GeV was set for any $\Delta m_+$ and 
$m_{\widetilde{\nu}}>300$ GeV. For smaller sneutrino masses, the sensitivity 
decreases due to the reduced pair production cross section and also due to 
reduced selection efficiency. In such situations where none of the above mass limits can apply, the generic lower 
limit of approximately 45 GeV, derived from the analysis of the $Z$-width, is still valid since this is independent of the field composition and of the 
decay 
modes of the charginos. Note that unlike neutralinos which couple to the $Z$-boson only via their higgsino component, the charginos couple to $Z$ via both 
their gaugino as well as higgsino components; so it is not possible to avoid the $Z$-width constraint for a light chargino.  

{\bf Sneutrino}: Light sneutrinos can only decay invisibly to $\nu \widetilde{\chi}^0_1$ unless the decays to charginos and heavier neutralinos are not 
kinematically suppressed. The invisible width of the $Z$-boson puts a lower limit on the left-sneutrino mass of 43.7 GeV, which improves slightly to 44.7 GeV if all three sneutrinos are mass-degenerate. Note that the lightest left-sneutrino by itself cannot be a cold DM candidate~\cite{falk, Hebbeker}, and we must introduce a mixing with a SM singlet sneutrino to make it a viable DM candidate (see, e.g.,~\cite{Arina:2007tm,rhs1,rhs2,rhs3,rhs4,rhs5,rhs6}). Since we are dealing here only with the MSSM field content and do not have a right-sneutrino component, we discard those points for which the sneutrino is the LSP. 

{\bf Slepton}: Studies of the $Z$-boson width and decays put a lower bound on the slepton masses $m_{\widetilde{\ell}_{R(L)}}>40~(41)$ GeV, independently of the decay modes for individual sleptons ($\widetilde{\ell}=\widetilde{e},~\widetilde{\mu},~\widetilde{\tau}$). This limit improves to 43 GeV if all the three slepton flavors are mass-degenerate. Tighter limits can be obtained assuming that 
sleptons are pair-produced at LEP, and each slepton dominantly decays to 
$\ell\widetilde{\chi}^0_1$, thus leading to two back-to-back leptons and missing transverse momentum. These limits are valid for a mass splitting $\Delta m_{\ell}=m_{\widetilde{\ell}}-m_{\widetilde{\chi}^0_1}>15$ GeV so that the final state leptons are not too soft. Moreover, the LEP results are interpreted assuming that only $\widetilde{\ell}_R\widetilde{\ell}_R$ production contributes, and hence, the limits are usually quoted for $\widetilde{\ell}_R$, since it is typically lighter than $\widetilde{\ell}_L$ in most SUSY models, and has a weaker coupling to the $Z$-boson so that the limits are more conservative. This is a good approximation for selectrons and smuons, but not for staus which can have significant mixing between 
the flavor eigenstates $\widetilde{\tau}_L$ and $\widetilde{\tau}_R$. The most 
conservative limit on the mass of the lightest stau is obtained with a mixing angle $\theta_{\widetilde{\tau}}\simeq 52^\circ$ which minimizes the production cross section.

The slepton mass limits of ${\cal O}(100)$ GeV quoted in Ref.~\cite{LEP-susy} 
 were derived under the assumption of gaugino mass unification at the GUT scale 
which was used to  fix the masses and composition of neutralinos. It was also 
assumed that the slepton branching ratio to $\ell\widetilde{\chi}^0_1$ is nearly 100\% which is a good approximation if the second lightest neutralino is heavy enough to suppress the cascade decay into $\ell\widetilde{\chi}^0_2$ followed by $\widetilde{\chi}^0_2\to f\bar{f}\widetilde{\chi}^0_1$ or $\widetilde{\chi}^0_1\gamma$. For smuons and staus, the LEP limits are independent of the MSSM parameters~\cite{Heister:2001nk}, and hence, directly applicable to our case. 
However, the selectron mass limit 
will be different if we do not assume gaugino mass unification due to a different production cross section involving the $t$-channel neutralino exchange, in addition to the usual $s$-channel $\gamma^*/Z$ exchange. Since there is no 
dedicated analysis of the LEP data addressing this issue available in the 
literature and this is beyond the scope of our present work, 
we only use the generic lower bound for $\widetilde{e}$ derived from the 
$Z$-width in our numerical analysis, but will also comment on the implications 
of the tighter selectron mass bound from LEP on our light neutralino scenario. 
We will also include the latest 95\% C.L. LHC exclusion limits for slepton pair production interpreted in the slepton-neutralino mass plane of pMSSM~\cite{ATLAS-slepton} which are applicable to both left- and right-handed 
selectrons and smuons.   

{\bf Gluino and Squarks}: The current LEP limits on the squark masses are 
similar to the slepton mass limits of ${\cal O}(100)$ GeV. However, since squarks are colored objects, their production cross sections are much higher at hadron colliders. The highest sensitivity on squark and gluino production now comes from the 
LHC experiments. The generic lower limit on the first/second generation squark masses is 600 - 750 GeV and on the gluino mass is 700 - 900 GeV, as set for 
simplified SUSY models by the ATLAS analysis of the 8 TeV LHC data~\cite{atlas-susy}. 
The corresponding CMS 
limits are very similar~\cite{cms-susy}. However, from the latest global fit 
of pMSSM after the LHC results~\cite{Hewett2} (see also Ref.~\cite{AbdusSalam:2009qd, AbdusSalam:2012ir} for an earlier global fit of pMSSM), the corresponding 
lower limit on squark mass is 
$m_{\widetilde{q}}\gsim 500$ GeV, and hence, we use this value to constrain our pMSSM parameter space.  

The LHC and Tevatron limits on the third generation squarks are usually weaker since the amount of bottom and top quark content in the proton is negligible, and hence, the direct production of bottom and top squark is suppressed with respect 
to the first/second generation squarks. The current exclusion limit for top 
squarks is between 300 - 600 GeV from the LHC data, depending on the decay channel~\cite{atlas-susy, cms-susy}. Similar limits have been derived for the bottom squark as well. Following the latest global fit of pMSSM~\cite{Hewett2}, we set the lower limits for light top squark and sbottom masses at 400 GeV and 300 GeV, 
respectively. Note that these limits are applicable as long as the light stop/sbottom is not highly mass degenerate with the lightest neutralino (with $\Delta m<10$ GeV or so) which always turns out to be the case for our light neutralino 
solutions satisfying all the other 
constraints. Thus we do not have any light sbottom solutions as 
considered in Ref.~\cite{Arbey:2012na}.  

    
The lower limits on the sparticle masses derived from the above discussion are 
summarized in Table~\ref{table2}. We emphasize here that for a light neutralino DM with $m_{\widetilde{\chi}^0_1}<m_Z/2$ as considered in our case, the dominant annihilation channels will be the $t$-channel processes mediated by light sfermions unless the $s$-channel $Z$-resonance or co-annihilation are effective. 
Hence, the lower limits on the chargino and sfermion masses as given in Table~\ref{table2} are crucial ingredients in our numerical analysis. 
\begin{table}[h!]
\begin{center}
\begin{tabular}{c|c|c}\hline\hline
Particle & Mass limit (GeV) & Validity Condition   
\\ \hline
 ${\widetilde{\chi}^\pm_1}$ & 103.5  & $m_{\widetilde{\chi}^+_1}-m_{\widetilde{\chi}^0_1}>3$ GeV, $m_{\widetilde{f}}>m_{\widetilde{\chi}^\pm}$\\
& 70 & $m_{\widetilde{\nu}}> 300$ GeV, $|\mu|\geq |M_2|$ \\
& 45 & generic LEP bound\\ \hline
${\widetilde{\mu}_R}$ & 88 & $m_{\widetilde{\mu}_R}-m_{\widetilde{\chi}^0_1} >15$ GeV, BR$(\widetilde{\mu}\to \mu\widetilde{\chi}^0_1)=1$ \\
${\widetilde{\tau}_1}$ & 76 & $m_{\widetilde{\tau}_1}-m_{\widetilde{\chi}^0_1} >15$ GeV, BR$(\widetilde{\tau}_1\to \tau\widetilde{\chi}^0_1)=1$ \\
${\widetilde{e}_R}$ & 95 & $m_{\widetilde{e}_R}-m_{\widetilde{\chi}^0_1} >15$ GeV, BR$(\widetilde{e}\to e\widetilde{\chi}^0_1)=1$, \\
& & $\mu=-200$ GeV, $\tan\beta=2$\\ \hline
$\widetilde{\ell}_{R~(L)}$ & 40~(41) & generic LEP bound\\
${\widetilde{\nu}}$   
& 43.7 & generic LEP bound\\ \hline
${\widetilde{g}}$ & 800 & \\
${\widetilde{q}}$ & 500 & \\ 
${\widetilde{t}_1}$ & 400 & \\
${\widetilde{b}_1}$ & 300 & \\
 \hline\hline
\end{tabular}
\end{center}
\caption{The lower limits on the sparticle masses used in our numerical 
analysis. The chargino and slepton mass limits are derived from the LEP 
data~\cite{LEP-susy} while the squark and gluino mass limits are derived from 
the LHC data~\cite{atlas-susy, cms-susy} which now 
supersede the LEP as well as the Tevatron~\cite{cdf, d0} limits.}
\label{table2}
\end{table} 
\subsection*{$\bm {B_s\to \mu^+\mu^-}$}
The rare decay $B_s\to \mu^+\mu^-$ is known to be a sensitive channel for new physics since its SM predicted rate is small due to helicity suppression. Recently, the first evidence for this decay was observed by the LHCb collaboration with the measured branching ratio: 
${\cal B}(B_s \to \mu^+\mu^-)=\left(3.2^{+1.5}_{-1.2}\right)\times 10^{-9}$~\cite{lhcb} which is in excellent agreement with the SM prediction of $(3.23\pm 0.27)\times 10^{-9}$~\cite{bsmu-SM}, thus raising some concerns for the ``health" of SUSY. However, it must be noted that the upper limit derived from the latest LHCb result is actually slightly {\it weaker} than the earlier upper limit of $<4.5\times 10^{-9}$~\cite{lhcb1}. The effect of the new results is mostly felt in the large $\tan\beta>50$ regions of the MSSM parameter 
space~\cite{Arbey:2012ax} which are also strongly constrained from direct SUSY searches and the MSSM Higgs searches. 
\subsection*{$\bm {b\to s\gamma}$}
We have also included the constraint from the radiative $B$-meson decay branching ratio, ${\cal B}(b\to s\gamma)=(3.55\pm 0.24\pm 0.09)\times 10^{-4}$~\cite{Asner:2010qj} which is somewhat higher than the SM prediction of $(3.15\pm 0.23)\times 10^{-4}$~\cite{Misiak:2006zs}. Thus large SUSY corrections are preferred which mainly occur for light chargino and top squarks and for large $\tan\beta$~\cite{Baer:1996kv}. 

\subsection*{$\bm {(g-2)_\mu}$}
Another important constraint comes from the muon anomalous magnetic moment $a_\mu\equiv (g-2)_\mu/2$ which gives a more than $3\sigma$ discrepancy with the SM prediction: $\delta a_\mu=(26.1\pm 8.0) \times10^{-10}$~\cite{Hagiwara:2011af}. 
The SUSY contribution to $\delta a_\mu$ can explain this discrepancy with relatively light smuons and/or large $\tan\beta$~\cite{g-2a, g-2b, g-2c}, and $\mu<0$ region of the SUSY parameter space is strongly disfavored (unless the electroweak gaugino masses $M_1,M_2<0$).
\subsection*{Higgs Sector}
For the light $CP$-even Higgs mass, we have chosen the value of $m_h=(125\pm 2)$ GeV, following the latest best-fit mass measurements of the Higgs-like particle discovered at the LHC: $(125.2\pm 0.3\pm 0.6)$ GeV (ATLAS)~\cite{atlas-dec12} and $(125.8\pm 0.4\pm 0.4)$ GeV (CMS)~\cite{cms-dec12}. For the other MSSM 
Higgs bosons, we ensure that all our allowed points satisfy the latest 
LHC constraints on the $m_A-\tan\beta$ plane~\cite{ATLAS-Atanb, CMS-Atanb} and on the MSSM 
charged Higgs mass~\cite{ATLAS-chargedh, CMS-chargedh} which are related to each other 
at tree-level by $m_{H^\pm}^2=m_A^2+m_W^2$. Note that the non-decoupling 
region with light $m_A\sim 95-130$ GeV, 
almost mass-degenerate with the light $CP$-even Higgs, and with the heavy 
$CP$-even Higgs SM-like, is disfavored~\cite{Christensen:2012ei, Carena:2013qia} by the latest LHC Higgs data and 
flavor constraints, especially the $B_s\to \mu^+\mu^-$ and $b\to s\gamma$. Hence, we work in the decoupling 
region with $m_A>300$ GeV with the light $CP$-even Higgs SM-like with mass in the vicinity of 125 GeV, and with the heavy $CP$-even Higgs nearly 
mass-degenerate with the $CP$-odd Higgs.  Also note that for a mostly bino-like neutralino, the $h\widetilde{\chi}^0_1\overline{\widetilde{\chi}}^0_1$ coupling is small enough to satisfy the current global limit on the Higgs invisible 
decay branching ratio: ${\rm BR}_{\rm inv}<0.28$ at 95\% CL~\cite{Giardino:2013bma}.    
\subsection*{Dark Matter Constraints}
The latest results from Planck give the current relic density of the cold dark 
matter content in our universe to be $\Omega_\chi h^2=0.1199\pm 0.0027$ at 68\% CL~\cite{Planck}. The corresponding value from the 9-year WMAP data is 
$\Omega_\chi h^2=0.1148\pm 0.0019$~\cite{wmap}. For the relic density of the neutralino DM in our case, we only require it to satisfy the WMAP $2\sigma$ upper bound combined with 10\% theoretical uncertainty: $\Omega_{\widetilde{\chi}^0_1} h^2 < 0.138$ which also encompasses the latest observed value from Planck. 
The cases where the 
neutralino relic density is below the corresponding WMAP lower bound, 
$\Omega_{\widetilde{\chi}^0_1}h^2<0.091$, could account for the correct 
relic density by alternative mechanisms of regeneration (see e.g.,~\cite{Hall:2009bx, Chu:2011be, Williams:2012pz}), or by invoking a multi-component DM 
scenario (see e.g.,~\cite{mdm1, mdm2, mdm3, mdm4, mdm5, mdm6, mdm7, mdm8, mdm9, mdm10}). 

As far as the DM direct detection constraints are concerned, since there is no unanimous upper bound on the direct detection cross section, we do not put this constraint a priori on the model parameter space. However, as we will see later, most of the allowed parameter space satisfying the other constraints also satisfy the most stringent upper limit on the spin-independent DM-nucleon scattering 
cross section set by the XENON100 experiment~\cite{Aprile:2012nq}. Note that for $m_\chi>10$ GeV, the spin-independent direct detection constraints are more 
stringent than the 
collider constraints from mono-jet~\cite{CMS-monoj, ATLAS-monoj} and isolated mono-photon~\cite{CMS-monoph, ATLAS-monoph} searches at the LHC  as well as from the LEP mono-photon data~\cite{LEP-mono}.  

Complementary to the direct detection constraints, there exist indirect detection constraints which are mostly sensitive to light WIMPs annihilating to SM fermions which eventually lead to gamma-ray signals. A lower limit of 
$m_{\widetilde{\chi}^0_1}\gsim 10$ GeV was derived from the CMB constraints~\cite{wmap-ind1, wmap-ind2, wmap-ind3, wmap-ind4, wmap-ind5, wmap-ind6}\footnote{Similar limits have also been derived from BBN constraints~\cite{bbn}.} for DM candidates with a velocity-independent annihilation cross section of $\langle \sigma_a v\rangle = 3\times 10^{-26}{\rm cm}^3\cdot {\rm s}^{-1}$.  Under the same assumption, the Fermi-LAT data put lower bounds of $m_{\widetilde{\chi}^0_1}\gsim 27$ GeV for annihilation to $b\bar{b}$ channel and $m_{\widetilde{\chi}^0_1}\gsim 37$ GeV for $\tau^+\tau^-$ channel~\cite{fermi-lat}. However, these bounds can be relaxed if we include the velocity-dependent contributions, as shown for pMSSM in Ref.~\cite{cotta}.  
In our numerical analysis, we include the latest 
Fermi-LAT 95\% CL upper limit on the integrated $\gamma$-ray flux from spectral line searches in the Milky Way galaxy: $\phi_\gamma < 4\times 10^{-10}{\rm cm}^{-2}\cdot{\rm s}^{-1}$~\cite{Ackermann:2012qk} for high latitude ($|b|>10^\circ$) plus a $20^\circ\times 20^\circ$ square at the galactic center and for 7 - 200 GeV energy range.

The various experimental constraints discussed above and used in our numerical analysis, in addition to the sparticle mass limits listed in Table~\ref{table2}, are summarized in Table~\ref{table3}. 
\begin{table}[!htb]
\begin{center}
\begin{tabular}{c|c}\hline\hline
Parameter & Constraint \\ 
\hline 
 $m_h$  & $(125\pm 2)$ GeV \\
$ \Gamma_Z^{\rm invisible}$ & $<3$ MeV \\ 
 $\Omega_{\widetilde{\chi}^0_1} h^2$ & $<0.138$ \\ 
$\phi_\gamma$ & $<4\times 10^{-10}~{\rm cm}^{-2}\cdot{\rm s}^{-1}$ \\
       $\mathcal{B}(B_s \rightarrow \mu^+\mu^-)$ & $ \left(3.2^{+1.5}_{-1.2}\right)\times 10^{-9}$ \\
       $\mathcal{B}(b \rightarrow s\gamma)$ &  $(3.55\pm 0.26)\times10^{-4}$ \\             $\delta a_\mu$ & $ (26.1\pm 8.0) \times10^{-10}$ 
\\ \hline\hline
\end{tabular}
\end{center}
\caption{The relevant experimental constraints used in our analysis, in addition to those listed in Table~\ref{table2}.}
\label{table3}
\end{table}
\section{Results}\label{sec4}
In order to scan the 19-dimensional pMSSM parameter space more efficiently while satisfying all the constraints listed in Tables~\ref{table2} and \ref{table3}, we have performed a MCMC analysis using a Gaussian 
distribution of likelihood function: $f(x,x_0,\sigma) = {\rm exp}[-(x-x_0)^2/2\sigma^2]$ for all the observables, with a preferred 
value $x_0\pm \sigma$. We have used \texttt{CaclHEP2.3}~\cite{calchep} and  
\texttt{micrOMEGAs2.4}~\cite{micromegas-1, micromegas-2, micromegas-3} 
to compute all the observables, together with \texttt{SoftSUSY}~\cite{Allanach:2001kg} for calculating the particle spectrum.  

\begin{figure}[tb]
\centering
\includegraphics[width=10cm]{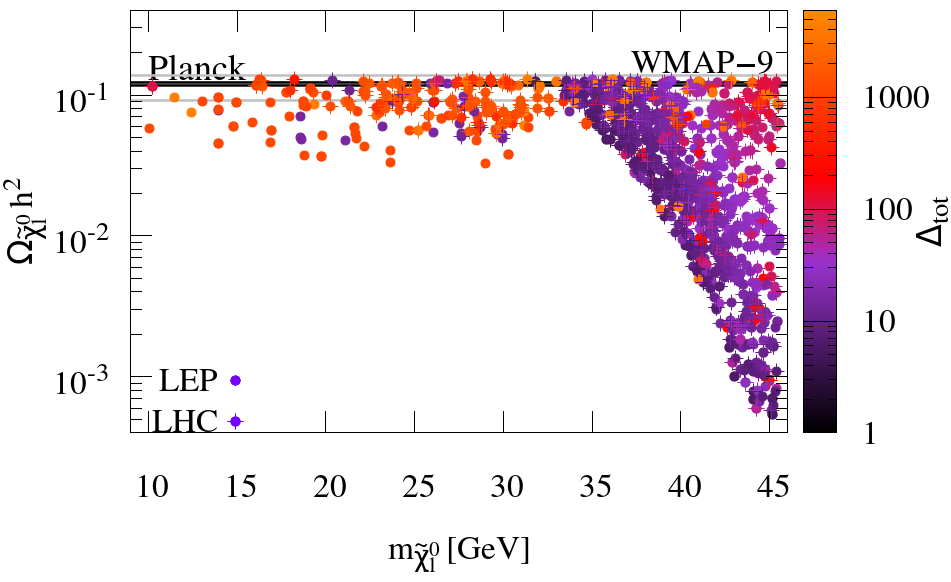}
\caption{The relic density of a light neutralino DM in pMSSM satisfying all the experimental constraints discussed in Section~\ref{sec3}. The color-coding denotes the fine-tuning measure defined by Eq.~(\ref{delta-tot}). The points denoted by circles satisfy all the experimental constraints, except that the squark 
masses are only required to satisfy the LEP lower limits. For the starred 
points (a subset of the circled points), the corresponding squark masses 
satisfy the latest LHC 
constraints. The top (bottom) grey horizontal line shows the $2\sigma$ upper (lower) limit of the cold dark matter relic density from WMAP-9 data, whereas the 
black (shaded) region shows the $1\sigma$ allowed range from the recent Planck data.}
\label{fig1}
\end{figure}

First we discuss our MCMC scan results for the relic density of a light 
neutralino DM candidate as shown in Figure~\ref{fig1}, which was obtained by numerically 
solving the Boltzmann equation using {\tt micrOMEGAS}~\cite{micromegas-1}. We require all the allowed points (shown as circles) to satisfy the experimental constraints given in Table~\ref{table3}, along with the LEP limits on sparticle masses given in Table~\ref{table2}. The latest LHC results put much tighter bounds on the strongly 
interacting squarks and gluinos and further eliminate some of these otherwise allowed parameter space, as shown by the starred points in Figure~\ref{fig1}. The WMAP-9 $2\sigma$ band is shown in grey, whereas the latest Planck result is shown as dark shaded region. We find that light neutralinos with mass as low as 10 GeV are still allowed, though severely fine-tuned with the electroweak fine-tuning measure defined by Eq.~(\ref{delta-tot}): $\Delta_{\rm tot}\gg 1$. This can be understood as follows 
by analyzing the gaugino and higgsino components of the lightest neutralino 
as well as its dominant annihilation channels.                         
\begin{figure}[tb]
\centering
\includegraphics[width=9cm]{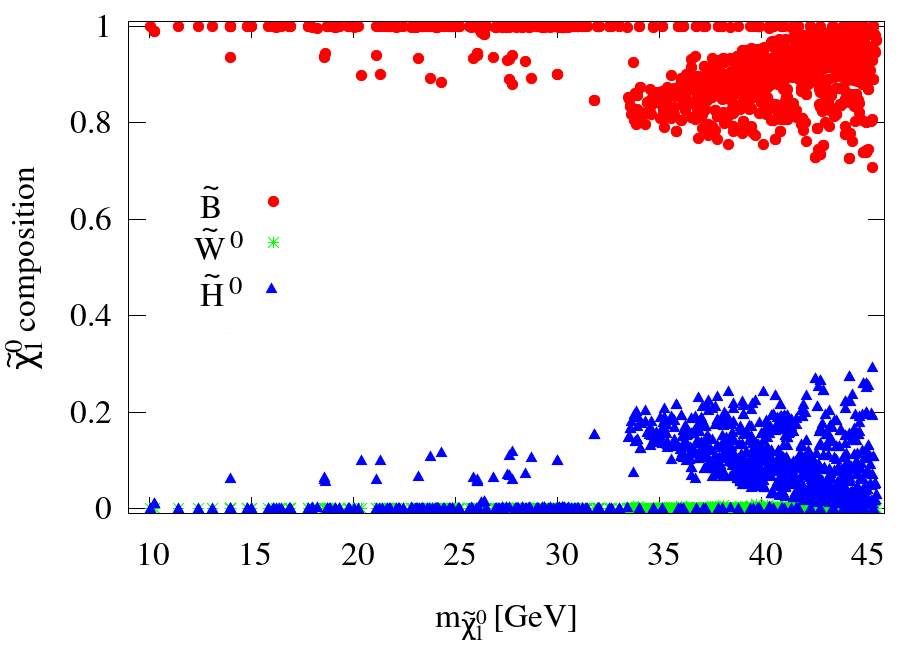}
\caption{The gaugino ($\widetilde{B},~\widetilde{W}^0$) and higgsino ($\widetilde{H}^0_d,~\widetilde{H}^0_u$) components of the lightest neutralino in our 
pMSSM parameter scan.}
\label{fig2}
\end{figure}

The bino, wino and higgsino fractions of the lightest neutralino for all the allowed points in our pMSSM parameter space are shown in Figure~\ref{fig2}. We reproduce the well-known result that the lightest neutralino is mostly bino-like 
for masses below $m_Z/2$, mainly due to the invisible $Z$-decay width 
constraint in Eq.~(\ref{gz}). However, a purely bino DM tends to overclose the 
universe unless it has an efficient annihilation channel to reach up to the 
thermal WIMP annihilation rate of $3\times 10^{-26}~{\rm cm}^3\cdot{\rm s}^{-1}$. One possibility is to consider a ``well-tempered" neutralino~\cite{ArkaniHamed:2006mb} which corresponds to the boundary between a pure bino and a pure higgsino or wino. Another possibility to reduce the bino relic density is by 
annihilation via the $t$-channel slepton exchange (the so-called ``bulk region") which is efficient for light sleptons, or by using co-annihilation with  a 
light slepton, squark, chargino or second-lightest neutralino in configurations where such light sparticles are not yet excluded by experimental searches. We find that most of 
the points with $m_{\widetilde{\chi}^0_1}$ close to 45 GeV can have either slepton co-annihilation or a resonant $Z$-annihilation due to a non-negligible higgsino component, and hence, can easily satisfy the WMAP upper limit on the 
relic density. These points are also less fine-tuned. On the other hand, the light neutralino DM points in the 10 - 30 GeV range 
as shown in Figure~\ref{fig1} have to be mostly bino-like and lie in the bulk region, thus leading to significant fine-tuning. Note that in the latter case, the next-to-lightest supersymmetric particle (NLSP) masses are 
much higher than the LSP mass, thus eliminating the possibility of a 
co-annihilation. 

\begin{figure}[tb]
\centering
\includegraphics[width=10cm]{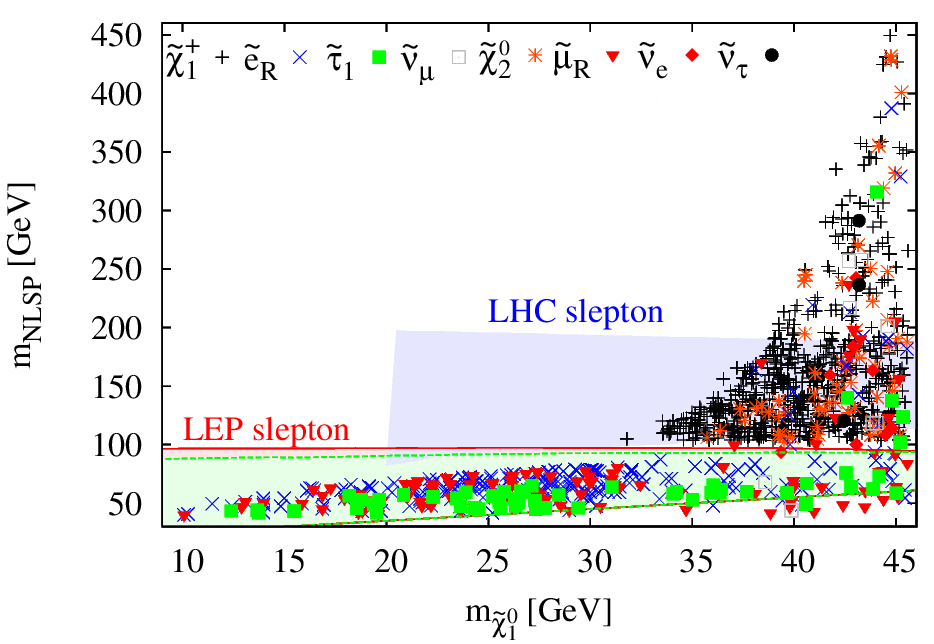}
\caption{The various NLSP masses as a function of the LSP mass for the allowed points (circles) shown in Figure~\ref{fig1}. The LEP exclusion regions strictly 
applicable for $\widetilde{\mu}_R$ (red shaded) and $\widetilde{\tau}_1$ (green shaded) and the LHC exclusion region for $\widetilde{\ell}~(l=e,\mu)$ (blue shaded) are also shown.}
\label{fig3}
\end{figure}

This is further clarified in Figure~\ref{fig3} where we show the various NLSPs and their masses as a function of the lightest neutralino mass. We see that all the allowed points with $m_{\widetilde{\chi}^0_1}<30$ GeV have a charged slepton NLSP 
with mass below 
100 GeV. Especially the points with a light stau are severely fine-tuned since they usually require a mass suppression by the off-diagonal elements in the slepton mass matrix, or a large $\mu$-term. We also show in Figure~\ref{fig3} the LEP exclusion regions in the charged slepton-neutralino mass plane, derived under the assumption of gaugino mass unification~\cite{LEP-susy}. The limits for 
light smuons and staus are still applicable to the pMSSM case as long as $\Delta m_\ell >15$ GeV, but not directly 
to light selectrons if we assume non-universal gaugino masses, and hence, can still allow the low neutralino mass regime. 
The latest 95\% C.L. ATLAS exclusion limits~\cite{ATLAS-slepton} are also 
shown in Figure~\ref{fig3} which were derived from searches for 
direct slepton (selectron and smuon) pair production and interpreted in the pMSSM. A similar dedicated analysis of the LEP data is 
required in order to {\it completely} rule out the light selectrons, and 
hence, the lightest neutralino DM mass below 30 GeV for the pMSSM scenario.

\begin{figure}[tb]
\centering
\includegraphics[width=7.3cm]{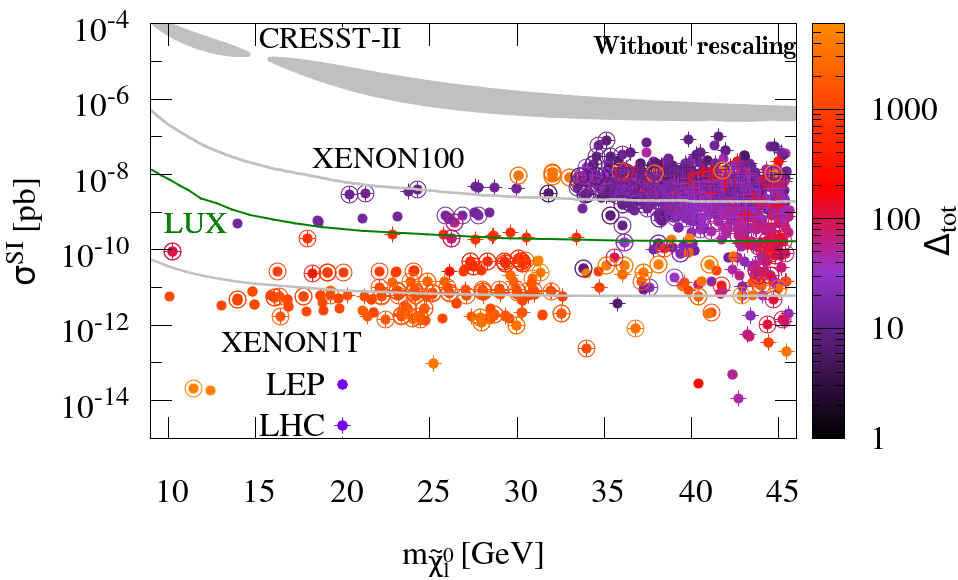}
\hspace{0.2cm}
\includegraphics[width=7.3cm]{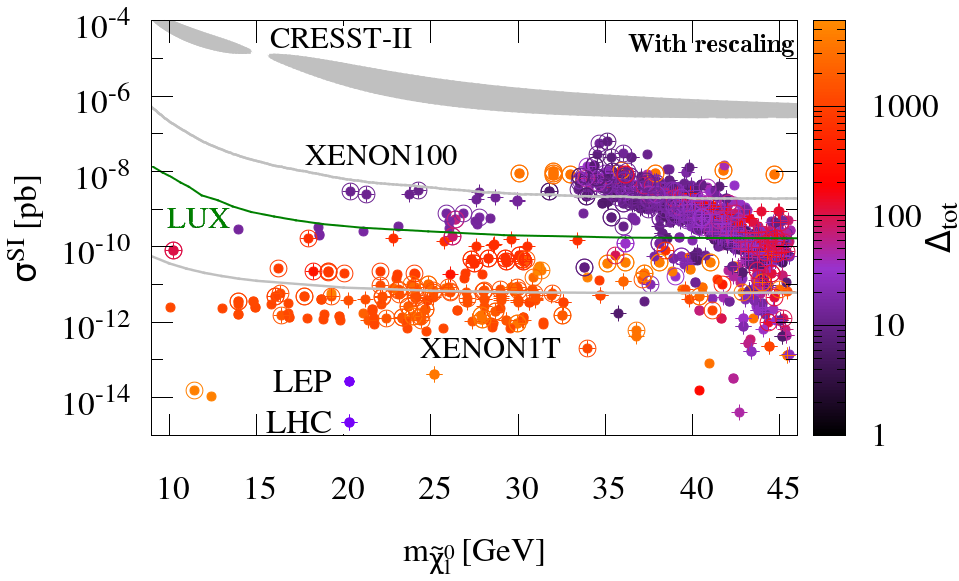}
\caption{The spin-independent direct detection cross-section values for the allowed points in our pMSSM scan. The color-coding and labeling of the points are same as in Figure~\ref{fig1}. The circled points correspond to those within the WMAP allowed band in Figure~\ref{fig1}. The left (right) panel shows the cross section without (with) rescaling of the DM density. The current upper limit from the XENON100 experiment and the projected limits from LUX and 
XENON1T experiments are shown as solid lines. The $2\sigma$-preferred range of 
CRESST-II is shown as the shaded region.}
\label{fig4}
\end{figure}

The neutralino DM-nucleon spin-independent scattering cross-sections for the allowed points is shown in Figure~\ref{fig4}. The points corresponding to the observed relic density within the WMAP band in Figure~\ref{fig1} are encircled. As mentioned earlier, the remaining points can also account for the correct relic 
density if we assume 100\% regeneration at late times 
(see e.g.,~\cite{Hall:2009bx, Chu:2011be, Williams:2012pz}). Another possibility is to invoke a multiple DM 
scenario (see e.g.,~\cite{mdm1, mdm2, mdm3, mdm4, mdm5, mdm6, mdm7, mdm8, mdm9, mdm10}) in which case the neutralino DM considered here will only constitute a fraction of the total observed DM density, and we must scale the neutralino 
DM density 
appropriately in order to calculate the neutralino-nucleon scattering cross 
section. Since the cross section depends linearly on the DM density, we use a 
rescaling factor of 
$r_\chi\equiv \Omega_{\widetilde{\chi}^0_1}/\Omega_{\rm observed}$, where for concreteness, we take the Planck central value for $\Omega_{\rm observed}h^2=0.12$. For the DM density distribution in the galactic halo, we have used the NFW profile~\cite{Navarro:1995iw, Navarro:1996gj}, as implemented in {\tt micrOMEGAs\_2.4}~\cite{micromegas-2}. The results with and without rescaling of the DM density are shown 
for comparison in 
Figure~\ref{fig4}. A comparison of the two panels in 
Figure~\ref{fig4} shows that the light neutralino solutions (with large fine-tuning) are mostly unaffected by the rescaling since these points yield a relic density value more than 10\% of the observed value (see Figure~\ref{fig1}). 
It is interesting to note that 
most of the allowed region in Figure~\ref{fig1} with low fine-tuning lead to a 
higher scattering cross section via $Z$-boson exchange and are 
already ruled out, even after rescaling, by the latest XENON100 data~\cite{Aprile:2012nq} or will be ruled out by 
the projected limit of XENON1T experiment~\cite{Aprile:2012zx} (and also LUX~\cite{LUX}) if they still get a null result. The points which survive with 
rescaling must be part of some multi-component DM scenario. In the light of the recent claims for positive hints of a light DM from some 
experiments, it is worth mentioning here that a few of our solutions with $\sigma^{\rm SI}\sim 10^{-7}$ pb are in the vicinity of 
the $2\sigma$ preferred range of the CRESST-II results~\cite{cresst}, but 
not compatible with the favored regions of DAMA~\cite{dama1, dama2} or CoGeNT~\cite{cogent1}. 

\begin{figure}[tb]
\centering
\includegraphics[width=7.3cm]{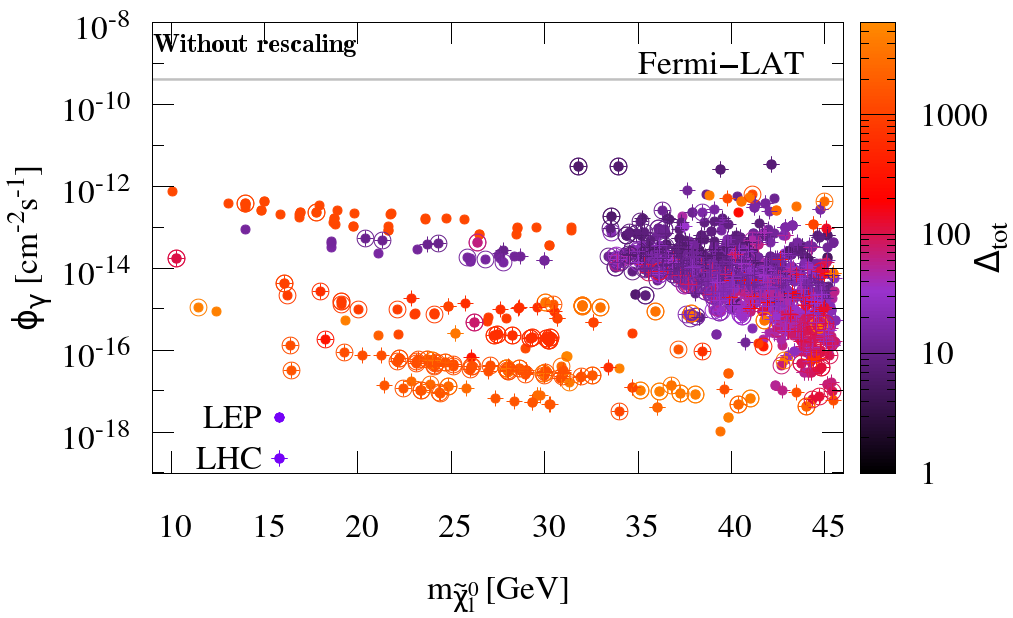}
\hspace{0.2cm}
\includegraphics[width=7.3cm]{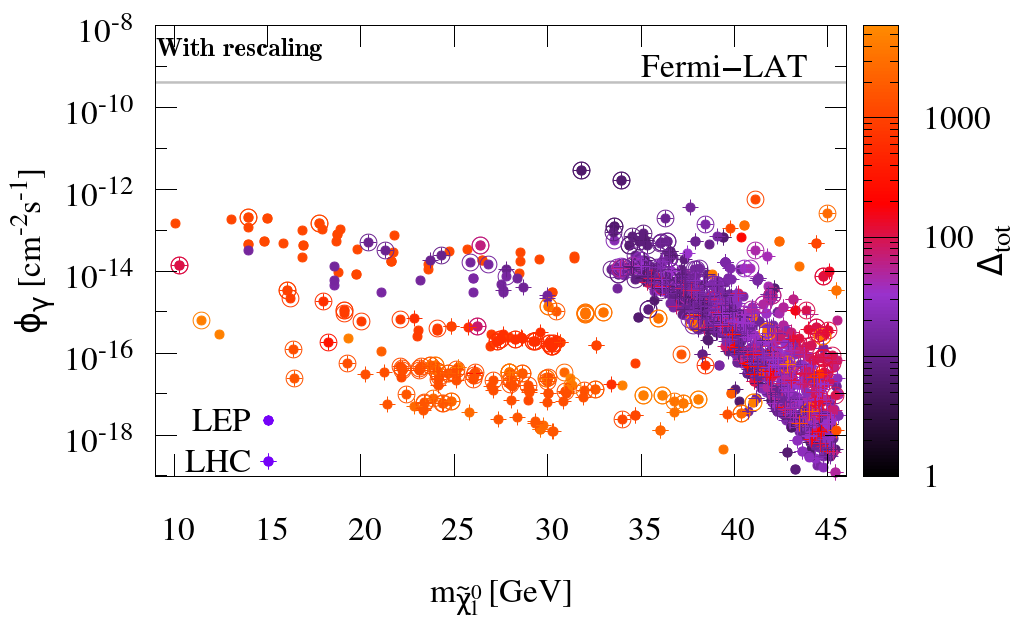}
\caption{The integrated photon flux from annihilation of the neutralino DM 
as a function of its mass. The color-coding and labeling of the points are same as in Figure~\ref{fig4}. The solid horizontal line shows the current upper 
limit from the Fermi-LAT data.}
\label{fig5}
\end{figure}

Figure~\ref{fig5} shows the integrated photon flux from spectral lines due to neutralino DM annihilation in the galactic halo. We have assumed the NFW profile for the DM density distribution, as implemented in {\tt micrOMEGAs\_2.4}~\cite{micromegas-3}. We do not consider  
galactic sub-structures or clumpy DM configurations since a proper analysis of these effects would require a detailed numerical simulation well beyond the scope of this work. As in Figure~\ref{fig4} for scattering cross-section, 
we have shown the fluxes for both the cases -- without and 
with scaling of the neutralino DM density. In the latter case, we have used the scaling factor $r_\chi^2$ for the thermally-averaged annihilation cross section $\langle \sigma_a v\rangle$ when the neutralino DM relic density is below the observed value. The light neutralino solutions with a relic density more than 10\% of the observed value (cf. Figure~\ref{fig1}) are mostly unaffected by the 
rescaling. Note that since the LSP in our case is mostly bino-like with heavy squarks and higgsinos and the charginos are not mass-degenerate with the LSP, the annihilation to photons is loop-suppressed, and hence, the photon line emission 
will be small. It is clear from Figure~\ref{fig5} that the current sensitivity of Fermi-LAT~\cite{fermi-lat} still leaves all of our allowed parameter space untouched. The future data from ongoing Fermi-LAT and next 
generation gamma-ray searches might be able to probe our allowed parameter 
space with a photon line signal~\cite{Sandick}.    

\begin{figure}[tb]
\centering
\includegraphics[width=7.3cm]{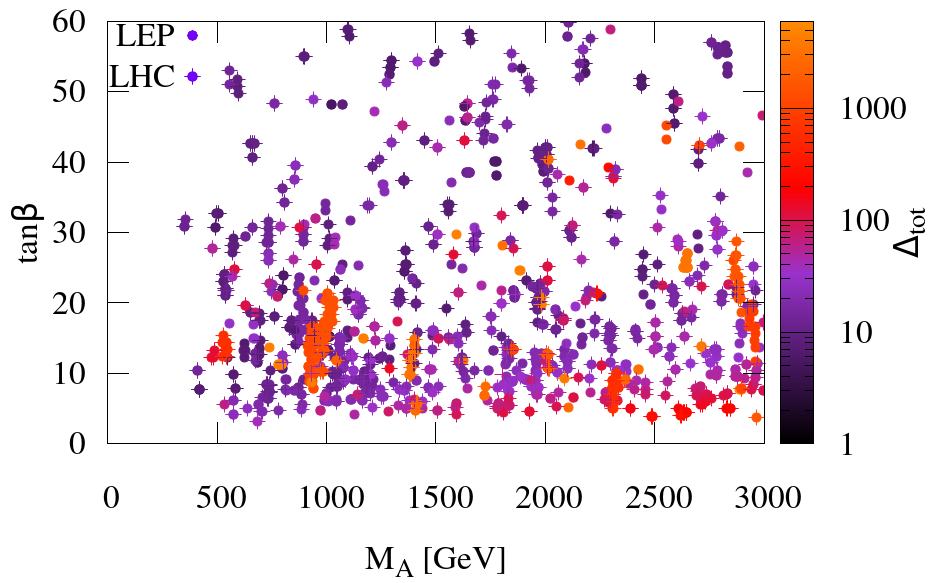}
\hspace{0.2cm}
\includegraphics[width=7.1cm]{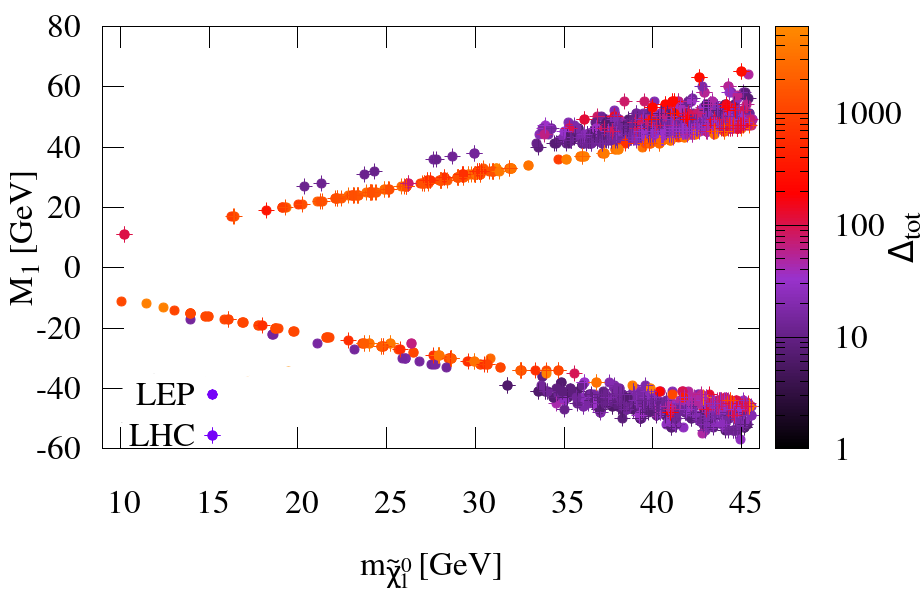}
\includegraphics[width=7.3cm]{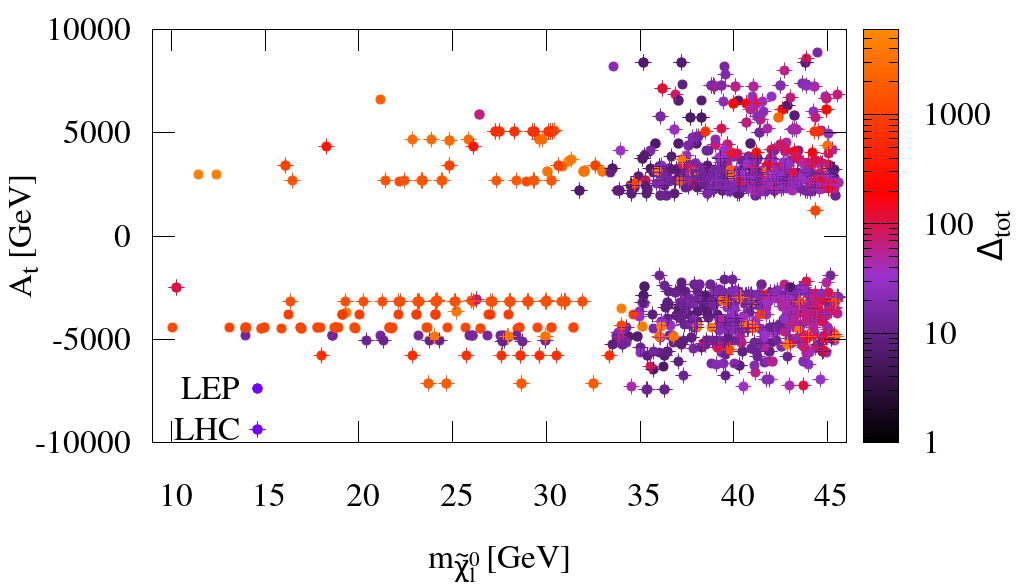}
\caption{The allowed parameter space for some of the relevant pMSSM parameters 
consistent with a light neutralino DM. The color-coding and labeling of the points are the same as in Figure~\ref{fig1}.}
\label{fig6}
\end{figure}

For completeness, we also show in Figure~\ref{fig6} some other relevant pMSSM 
parameters with respect to the lightest neutralino mass. The $m_A-\tan\beta$ 
parameter space is mostly consistent with the latest MSSM Higgs sector limits 
from the LHC~\cite{ATLAS-Atanb, CMS-Atanb}. As for the bino mass parameter $M_1$, it is clear that a relatively small value of $|M_1|<100$ GeV is preferred to obtain a light bino-like neutralino LSP. Finally, as is well-known, a relatively large value of $|A_t|$ is required in order to enhance the radiative corrections for the light $CP$-even Higgs mass to be consistent with the LHC-preferred value of $125\pm 2$ GeV.  

Finally, we wish to point out that the allowed sparticle spectra 
discussed 
here with mostly heavy squarks, and with light sleptons, chargino and bino-like neutralino LSP, are also crucial for explaining the muon $(g-2)$ anomaly~\cite{endo}, while simultaneously satisfying all the other experimental constraints. Light staus with sizable left-right mixing can also lead to an enhanced 
$h\to \gamma\gamma$ decay rate~\cite{Carena:2012gp, Carena:2013iba, 
Kitahara:2013lfa}, which might be able to explain the persistent excess in the Higgs signal strength in this channel: $1.65\pm 0.24({\rm stat})^{+0.25}_{-0.18}({\rm syst})$~\cite{atlas-gamma}. 
\section{Conclusion}\label{sec5}
We have studied the naturalness of a light neutralino dark matter candidate in the MSSM in the light of the latest results from the collider (LEP, Tevatron and LHC), flavor (LHCb) and dark matter (XENON100, WMAP-9, Planck) 
sectors. In particular, keeping in mind the recent positive hints for a light DM below 45 GeV in some direct detection experiments and the null results from SUSY searches at the LHC, we perform a dedicated study focusing on the possibility of a light DM candidate in the form of the lightest neutralino in the pMSSM, also analyzing the naturalness of such a scenario. We include all the new results from the ongoing LHC experiments in our analysis, such as the null results from the SUSY searches, the observation of a Higgs-like particle, the observation of the rare meson decay $B_s\to \mu^+\mu^-$, and the updated constraints on the MSSM Higgs sector. We also take into account the 9-year WMAP results as well as the recently released Planck results for the DM relic density and the Fermi-LAT data for the integrated photon flux. In order to efficiently scan over the 19-dimensional pMSSM parameter space, we perform a MCMC likelihood analysis focusing on a light neutralino with mass below the LEP bound of 46 GeV (applicable to gaugino mass unification models). 

We find that a light neutralino DM with mass as low as 10 GeV is still allowed in the pMSSM, while satisfying all the existing experimental constraints. However, such neutralinos which are required to be mostly bino-like are severely fine-tuned and require the existence of light sleptons with mass below 100 GeV in order to provide an efficient annihilation channel to reduce the DM relic density below the observed upper limit. Such light smuons and staus are excluded from LEP searches while light selectrons are excluded only if we assume gaugino mass unification. A dedicated analysis of the LEP data in the context of a pMSSM scenario could completely eliminate the possibility of a light neutralino DM in the 
mass range of 10 - 30 GeV. We also find that for the allowed parameter space, 
other possible solutions in this mass range as discussed in the literature (e.g., light sbottom NLSP) are now excluded mainly due to the 
latest LHC results on the strongly interacting sfermion sector.  

On the other hand, low fine-tuning regions can be obtained around $m_{\widetilde{\chi}^0_1}=45$ GeV where the resonant annihilation via the $s$-channel 
$Z$-exchange is possible for a bino-higgsino mixture of neutralino LSP. However, such regions also predict a higher spin-independent DM-nucleon scattering cross section, and are mostly excluded by the recent XENON100 limits. The remaining 
such points are within the reach of the XENON1T projected limits.

In conclusion, while a light neutralino DM with mass below 46 GeV has been conclusively ruled out in MSSM with gaugino mass unification by LEP searches, such a possibility in a general version of MSSM is of enormous interest in the light of the recent claims for a positive signal in some DM direct detection experiments. Taking into account the latest experimental results from collider, flavor, dark 
matter and astrophysical/cosmological sectors, we show that such a light 
neutralino DM scenario is also getting highly constrained in MSSM without 
gaugino mass unification. However, within a pMSSM scenario, 
there still exists some parameter space for light neutralino DM which could be completely probed by a dedicated analysis of the existing experimental data, 
in combination with the ongoing searches.   
\section*{Acknowledgments} 
We thank Matthew Dolan and Patrick Janot for valuable discussions and comments on the draft, and Jonathan Da Silva for useful discussions regarding the 
MCMC analysis. CB is supported by the ERC advanced grant `DARK' at IAP, Paris. 
The work of PSBD and AM is supported by the 
Lancaster-Manchester-Sheffield Consortium for Fundamental Physics under STFC 
grant ST/J000418/1. EP is supported by STFC ST/J501074.
\section*{Note Added}
After the submission of our paper, another DM direct detection experiment, namely, CDMS-II~\cite{Agnese:2013rvf}, has reported three WIMP-candidate events with an expected background of 0.7 events. Their best-fit WIMP mass is 8.6 GeV with the WIMP-nucleon cross section of $1.9\times 10^{-5}$ pb. This bolsters our motivation in this work to examine the allowed parameter space for a 
light neutralino DM candidate in the MSSM.

\end{document}